\documentclass[11pt,a4paper]{article}
\usepackage[nohyperref]{emnlp-ijcnlp-2019}
\usepackage{times}
\usepackage{latexsym}

\usepackage{color}
\usepackage{float}
\usepackage{tabularx}
\usepackage{subfig}
\usepackage{multirow}
\usepackage{latexsym}
\usepackage{amsmath}
\usepackage{amssymb}
\usepackage{colortbl}
\usepackage{array}
\usepackage{bm}
\usepackage{booktabs}
\usepackage{pgfplots}

\usepackage{algorithm, algorithmic}

\newcommand{\Vh}{\mathbf{h}}

\newcommand{\Vk}{\mathbf{k}}
\newcommand{\Vv}{\mathbf{v}}

\newcommand{\Vb}{\mathbf{b}}

\newcommand{\Vq}{\mathbf{q}}

\newcommand{\MW}{\mathbf{W}}

\newcommand{\MQ}{\mathbf{Q}}
\newcommand{\MK}{\mathbf{K}}
\newcommand{\MV}{\mathbf{V}}
\newcommand{\MH}{\mathbf{H}}

\newcommand{\xdownarrow}[1]{%
  {\left\downarrow\vbox to #1{}\right.\kern-\nulldelimiterspace}
}

\DeclareMathOperator{\softmax}{softmax}
\DeclareMathOperator{\GraphAttention}{GraphAtt}
\DeclareMathOperator{\Attention}{Attention}
\DeclareMathOperator{\Norm}{Norm}
\DeclareMathOperator{\FFN}{FFN}

\usepackage{url}

\aclfinalcopy 

\newcommand\dsname{SciTSR}
\newcommand\dshard{SciTSR-COMP}
\newcommand\modelname{GraphTSR}

\title{Complicated Table Structure Recognition}

\author{Zewen Chi, Heyan Huang, Heng-Da Xu, Houjin Yu, Wanxuan Yin, Xian-Ling Mao \\
  Department of Computer Science and Technology, \\
Beijing Institute of Technology, China \\ 
  {\tt czwin32768@gmail.com}}

\date{}

\begin{document}
\maketitle
\begin{abstract}
The task of table structure recognition aims to recognize the internal structure of a table, which is a key step to make machines understand tables.
Currently, there are lots of studies on this task for different file formats such as ASCII text and HTML. It also attracts lots of attention to recognize the table structures in PDF files.
However, it is hard for the existing methods to accurately recognize the structure of complicated tables in PDF files. The complicated tables contain spanning cells which occupy at least two columns or rows.
To address the issue, we propose a novel graph neural network for recognizing the table structure in PDF files, named \modelname{}.
Specifically, it takes table cells as input, and then recognizes the table structures by predicting relations among cells.
Moreover, to evaluate the task better, we construct a large-scale table structure recognition dataset from scientific papers, named \dsname{}, which contains 15,000 tables from PDF files and their corresponding structure labels. Extensive experiments demonstrate that our proposed model is highly effective for complicated tables and outperforms state-of-the-art baselines over a benchmark dataset and our new constructed dataset.\footnote{\url{https://github.com/Academic-Hammer/SciTSR}}
\end{abstract}

\section{Introduction}

The task of table structure recognition
is to recognize the internal structure of a table, which is a key step to make machines understand tables.
The recognized machine-understandable tables have many potential applications including question answering, dialogue systems and table-to-text \cite{pasupat2015compositional, jauhar2016tablesqa, li2016deep, yan2016docchat, bao2018table, jain2018tablesum}.

\begin{figure}[t]
\begin{center} 
\includegraphics[width=0.8\linewidth]{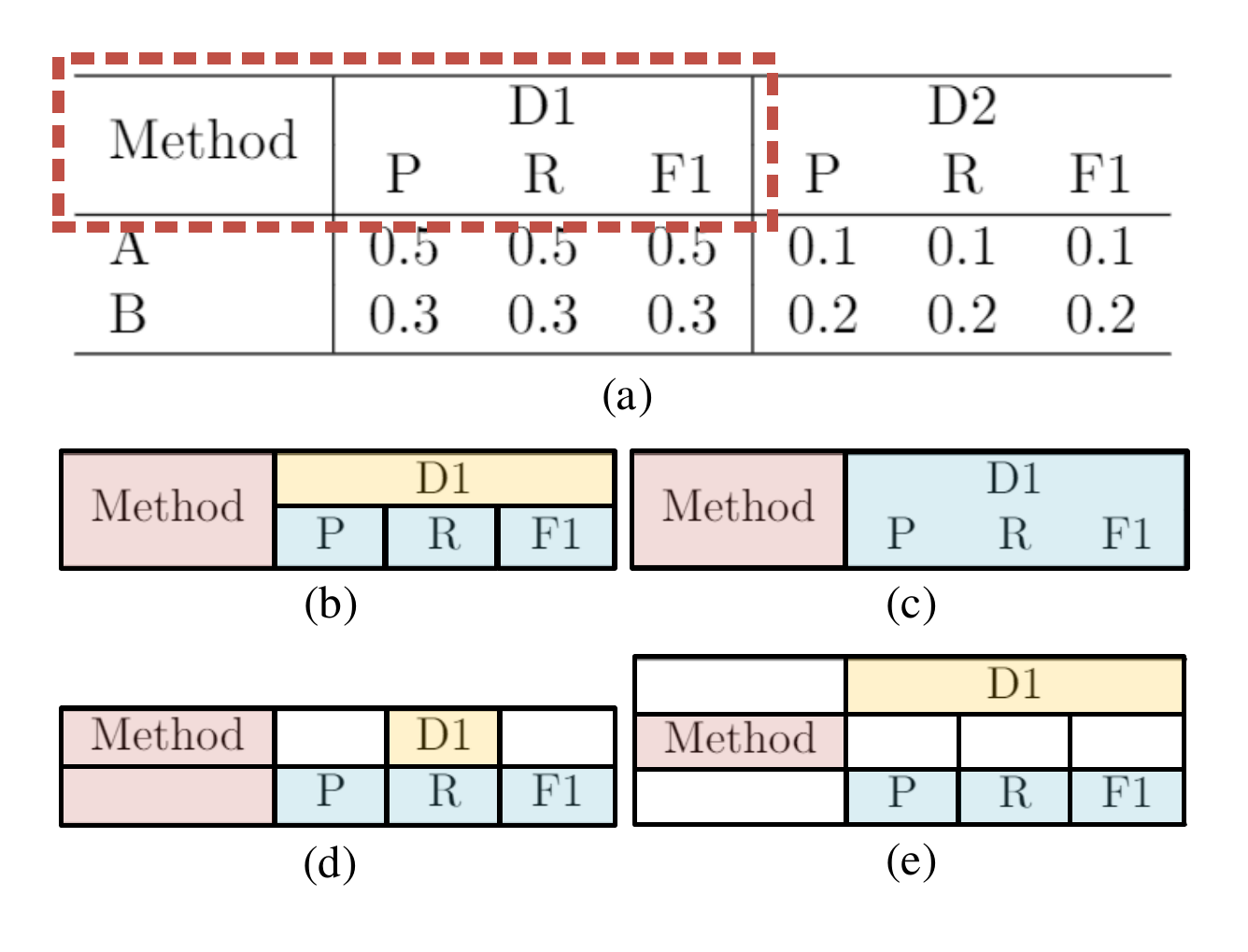}
\caption{An intuitive example of a complicated table with spanning cells. The example table is shown in (a), and (b) is the real structure of the dashed box area.
The recognized structure by existing methods are shown in (c) - (e). Note that in (c), the four cells on the right side are incorrectly recognized as a single cell.
} 
\label{fig:example}
\end{center} 
\end{figure}

Currently, table structure recognition has been studied for many file formats such as ASCII text, HTML and image.
As a popular and widely-used file format, PDF also attracts lots of attention, and many approaches have been proposed for PDF format.
The existing approaches can be classified into two categories: rule-based methods \cite{ramel2003detection, yildiz2005pdf2table, hassan2007table} and data-driven methods \cite{schreiber2017deepdesrt, li2019tablebank}.

However, it is hard for the existing methods to accurately recognize the structure of complicated tables in PDF files.
A spanning cell is a table cell that occupies at least two columns or rows.
If a table contains spanning cells, it is called a complicated table.
We provide an intuitive example of a complicated table in Figure \ref{fig:example}
, and we show a table in a PDF file in Figure \ref{fig:example} (a). 
In Figure \ref{fig:example} (b), we show the real structure of the table area enclosed by the dashed box in Figure \ref{fig:example} (a).
In Figure \ref{fig:example} (c) - (e), we present the recognized structure by existing methods, i.e., Adobe Arcobat SDK, DeepDeSRT \cite{schreiber2017deepdesrt} and Tabby \cite{shigarov2016configurable}.
It can be observed that none of these existing methods provides a correct result in this case.
Although the proportion of spanning cells is usually small in a complicated table, they contain more important semantic information than other cells, because they are more likely to be table headers in a table.
The table header of a table is crucial to understand the table.
Therefore, the recognition of complicated table structures is an important problem to be solved.

To address the aforementioned issue, we propose a novel graph neural network model that reformulates the task as a edge prediction problem on graphs.
Specifically, it encodes a table by a stack of graph attention blocks, and then recognizes the table structure by predicting relations among cells.
Additionally, because there is no available training data for this task, we construct a new large-scale dataset for table structure recognition in PDF files, in which tables are collected from scientific papers, denoted as \dsname{}.
It contains 15,000 tables and the corresponding structure labels.
Over a benchmark and our \dsname{} dataset, extensive experiments demonstrate that our proposed model  outperforms state-of-the-art baselines greatly, especially in the case of complicated tables.
Our contributions are two-fold:

\begin{itemize}
    \item We
     propose a novel graph neural network model to recognize the structure of tables in PDF files, especially complicated tables. Extensive experiments demonstrate that our proposed model outperforms state-of-the-art baselines greatly.
    \item We
      construct a new large-scale table structure recognition dataset from scientific papers, which contains 15,000 tables and corresponding structure labels. To the best of our knowledge, this is the first large-scale dataset for table structure recognition in PDF files.
\end{itemize}

\begin{figure*}[th]
\begin{center} 
\includegraphics[width=1.0\linewidth]{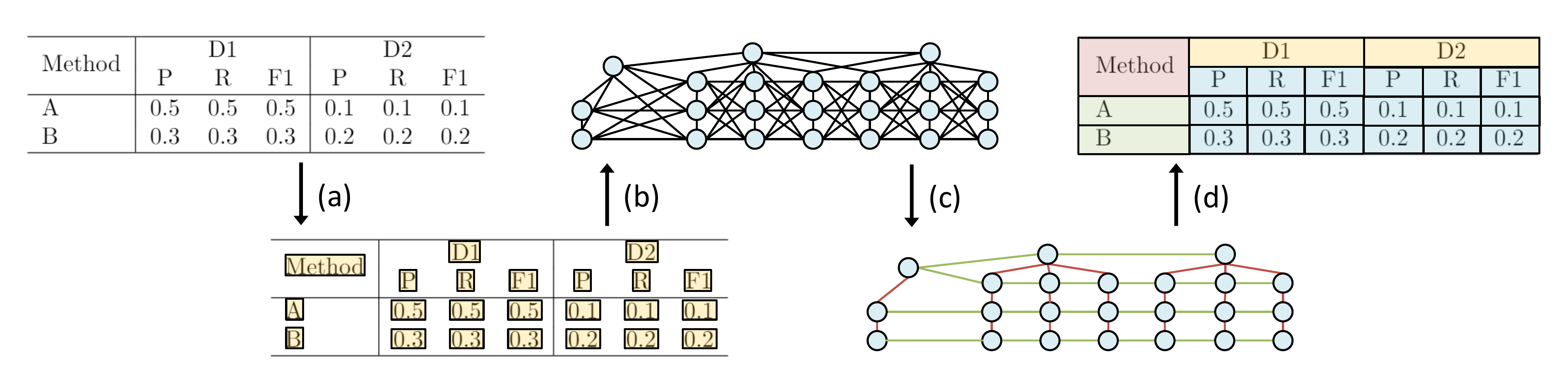}
\caption{Overview of our method. Given a table in PDF as input, our method recognize its structure by the following four steps: (a) Pre-processing: obtaining cell contents and their corresponding bounding box from PDF; (b) Graph construction: building an undirected graph on these cells; (c) Relation prediction: predicting adjacent relations by our proposed \modelname{}; (d) Post-processing: recovering table structure from the labeled graph.
} 
\label{fig:ov}
\end{center} 
\end{figure*}

\section{Related Work}

In this section, we will introduce the related methods and datasets on the task of able structure recognition in PDF files. 

\subsection{Methods}

Existing methods can be classified into two categories: rule-based methods and data-driven methods.

\paragraph{Rule-based methods}
In 2003, \citeauthor{ramel2003detection} utilized ruling lines and arrangement of text components to recognize table structures from exchange format files \cite{ramel2003detection}. 
Later, \citeauthor{yildiz2005pdf2table} proposed to recognize columns by computing horizontal overlaps of texts, by using the text blocks generated by \textit{pdftohtml}\footnote{\url{http://pdftohtml.sourceforge.net/}} tool as input \cite{yildiz2005pdf2table}.
In 2007, \citeauthor{hassan2007table} employed a similar idea to the work of \citet{ramel2003detection} but extended it to PDF format.
In 2009, \citeauthor{oro2009trex} proposed to recognize table structures by grouping basic content elements in a bottom-up way \cite{oro2009trex}.
Unlike these methods that use ``hard'' rules, \citeauthor{shigarov2016configurable} presented a ``soft'' rule-based method that can be adapted to different domains \cite{shigarov2016configurable}.
 
\paragraph{Data-driven methods}

As far as we know, there are two data-driven methods that leverage deep learning to solve this task.
In 2017, \citeauthor{schreiber2017deepdesrt} first proposed a model called DeepDeSRT, which treats the table structure recognition task as an image semantic segmentation problem, and then recognizes the regions of columns and rows respectively \cite{schreiber2017deepdesrt}.
In 2019, \citeauthor{li2019tablebank} proposed an image-to-text model, which is first trained to encode the table image and then decode the table structure as a HTML-like sequence of tags \cite{li2019tablebank}.
Despite the tag sequence is designed to represent a table, it doesn't provide column coordinates of cells. 
Thus, the tag sequence cannot be used to restore a table, which means their model is not a complete model for this task.
So it cannot be used as a baseline in our experiments.

Overall, all these approaches only work well on simple grid-like tables but fail on the complicated tables with spanning cells. Therefore, we propose a novel graph neural network model to address this issue, and it will be introduced in section 3.

\subsection{Datasets}
There are two related datasets for table structure recognition. One is the ICDAR-2013 dataset from ICDAR 2013 Table Competition \cite{gobel2013icdar}, which has only 156 tables in total. Although every PDF document is well labeled, the size of the dataset is too small to support data-driven models. Recently, \citet{li2019tablebank} releases a large-scale dataset called TableBank for table detection and structure recognition.
However, TableBank only provides tables in image format, and column coordinates information is missed in its structure labels. 
That means it cannot be used for this task.
So we construct a new large-scale table structure recognition dataset from scientific papers for this task, which will be introduced in section 4.

\section{Method}

Figure \ref{fig:ov} illustrates an overview of our method. 
Given a table in PDF format as input, our method recognizes its structure by the following four steps: (a) Pre-processing: obtaining cell contents and their corresponding bounding box from PDF; (b) Graph construction: building an undirected graph on these cells; (c) Relation prediction: predicting adjacent relations by our proposed \modelname{}; (d) Post-processing: recovering table structure from the labeled graph.
We use the same pre-processing procedure as \cite{shigarov2016configurable}, and employ a simple post-processing to convert a labeled graph to structure data.
Thus, in this section, we mainly focus on introducing step (b) and (c).
We first formally define the problem in section 3.1, and then introduce step (b) and (c) in section 3.2 and 3.3.

\subsection{Problem Definition}


Consider that each cell in a table can be viewed as a vertex, and an adjacent relation (i.e., vertical, horizontal) can be viewed as a labeled edge. So a table can be represented as a graph with labeled edges $T = \left \langle  V, R \right \rangle$, where $V$ is a set of vertices, and $R \subseteq V \times V \times \{vertical, horizontal\}$ is a set of relations. The relation set $R$ is actually the table structure we want to recognize.
Given a set of vertices $V$ of table $T$ as input, the problem is to find out an approximation to the real relations $R$.

\subsection{Graph Construction}

Let $E \subseteq V \times V$ denote the unlabeled edges of $R$, the goal of graph construction is to build such edges $E'$ so that $|E' \cap E|$ is as large as possible.
The simplest idea is to construct a complete graph where $E'=V \times V$. However, it is impractical because it requires our model to encode $O(|V|^2)$ edges, and thus we need to reduce $|E'|$ in a reasonable size.
In this paper, we use a simple $K$ nearest neighbors ($K$NN) method to construct $E'$, in which each vertex is connected to its $K$ nearest neighbors so that the total number of edges will be reduced to $O(K|V|)$.
In the next sub-section, we will introduce \modelname{}, which takes unlabeled $E'$ as input and classifies each edge into one of the categories of \textit{``vertical''}, \textit{``horizontal''} and \textit{``no relation''}.

\begin{figure}[th]
\begin{center} 
\includegraphics[width=1.0\linewidth]{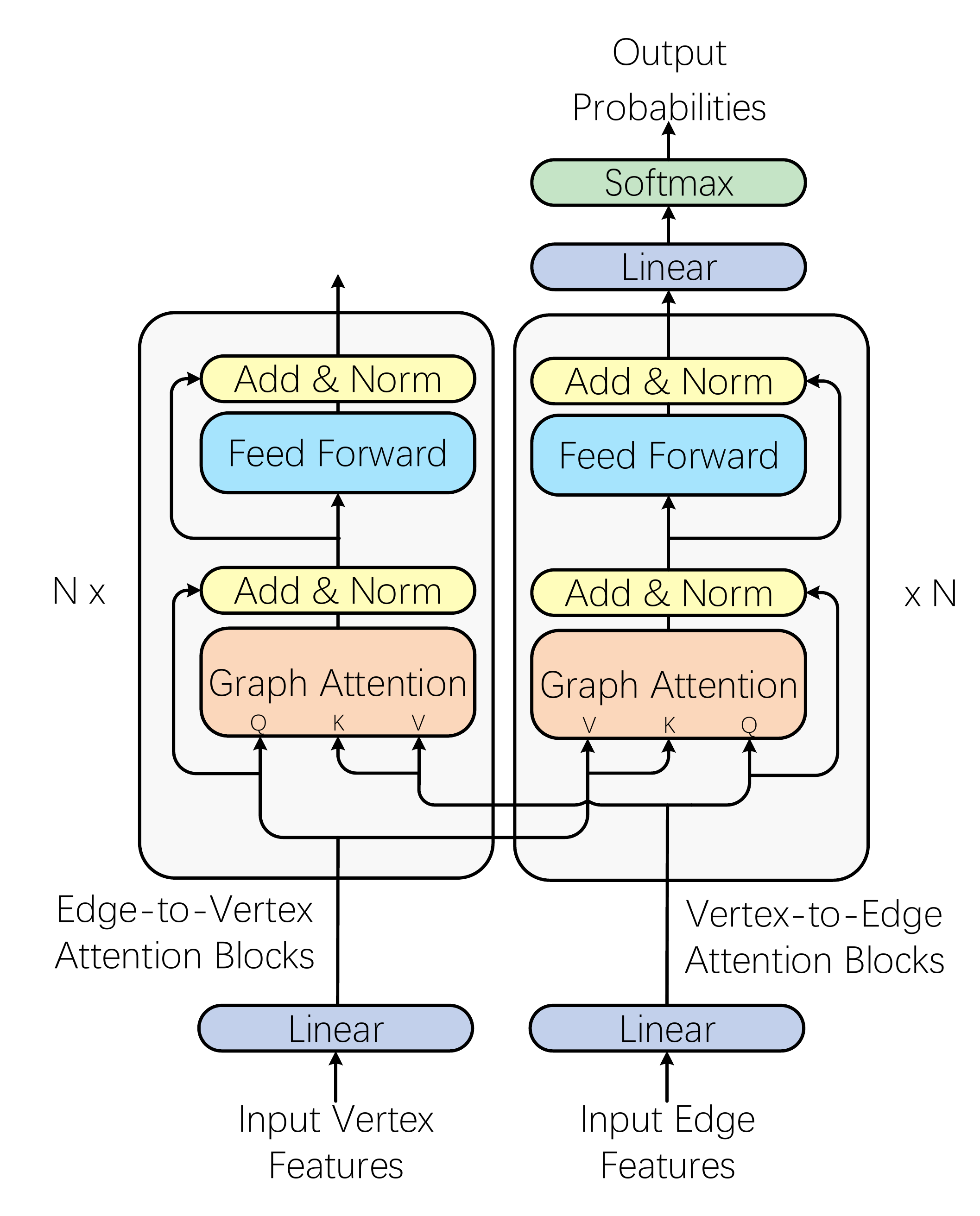}
\caption{The architecture of our proposed \modelname{}.
} 
\label{fig:model}
\end{center} 
\end{figure}

\subsection{\modelname{}}

Our proposed \modelname{} is illustrated in Figure \ref{fig:model}.
It takes the vertex and edge features of a graph as input, and then computes their representations by $N$ edge-to-vertex graph attention blocks and $N$ vertex-to-edge graph attention blocks, respectively.
Finally, it performs a classification over these edges.

\paragraph{Vertex and edge features} We design several features as the initial representation of
vertices and edges. Three types of vertex features are designed including the size of cells,
the absolute locations and the relative locations.
As for the edge features, we use several measures of distance between cells including
Euclidean distance, x-axis distance and y-axis distance in both absolute and relative manner.
We also compute the overlap of cell pairs along x-axis and y-axis as the edge features.

\paragraph{Graph attention}
Let us first review the commonly-used scaled dot-product attention proposed by \citet{vaswani2017attention}:
$$\Attention(\MQ, \MK, \MV) = \softmax(\frac{\MQ\MK^\top}{\sqrt{d_k}})\MV$$
where $d_k$ is the dimension of keys, and $\MQ, \MK, \MV$ are the matrices of ``queries'', ``keys'' and ``values'', respectively.
Differently, inspired by the work of \citet{velivckovic2017graphattention}, the graph attention of our proposed \modelname{} doesn't draw global dependencies between nodes but local dependencies on neighboring nodes, which is computed as:
$$a_{iu} = \frac{e^{\Vk_i^\top \Vq_u/\sqrt{d_k}}}{\sum_{j \in \mathcal{N}(u)} e^{\Vk_j^\top \Vq_u/\sqrt{d_k}} }$$
where $\Vk_i, \Vq_u \in \mathbb{R}^{d_\Vk}$ are the $i^\text{th}$ and $u^\text{th}$ row of $\MK$ and $\MQ$, and
$\mathcal{N}(u)$ is the set of neighbors of node $u$.
Let $\Vv_j \in \mathbb{R}^{d_v}$ be the $j^\text{th}$ row of $\MV$. 
Then the graph attention function is:
$$\GraphAttention(\Vq_u, \MK, \MV) = \sum_{j \in \mathcal{N}(u)} a_{ju}\Vv_j$$

\paragraph{Graph attention blocks}
To support edge features, the original graph is converted to a bipartite graph with additional nodes that represent edges in the original graph.
With this setting, we use $N$ edge-to-vertex graph attention blocks and $N$ vertex-to-edge graph attention blocks to encode vertices and edges separately.
As shown in Figure \ref{fig:model}, the computation of the two kinds of attention blocks is symmetrical, so we only introduce the computation of an edge-to-vertex attention block.

In the \modelname{}, we simply set $d_k = d_v = d$, which means we use the same dimension size for ``keys'' and ``values''.
Suppose that the outputs of the $(n-1)^\text{th}$ edge-to-vertex and vertex-to-edge attention block are $\MH_v^{(n-1)} \in \mathbb{R}^{|V| \times d}$ and $\MH_e^{(n-1)} \in \mathbb{R}^{|E| \times d}$, respectively.
We denote them as $\MH_v$ and $\MH_e$ for simplicity.
At the $n^\text{th}$ edge-to-vertex attention block, we first compute $\MQ, \MK, \MV$ by:
$$\MQ = \MH_v\MW_Q, \MK = \MH_e\MW_K, \MV = \MH_e\MW_V$$
where $\MW_Q,\MW_K,\MW_V \in \mathbb{R}^{d \times d}$ are learnable parameters.
That means the edge nodes serve as ``keys'' and ``values'', in the edge-to-vertex attention block.
Then, by attending neighboring edge nodes, the representation of vertex nodes is updated as:
$$ \tilde{\MH}_v^{(n)} = \Norm(\MH_v + \GraphAttention(\MQ, \MK, \MV))$$
where $\Norm$ is the layer normalization function \cite{lei2016layer}.
Finally, the output of the $n^\text{th}$ attention block is calculated as:
$$\MH_v^{(n)}=\Norm(\tilde{\MH}_v^{(n)} + \FFN(\tilde{\MH}_v^{(n)}))$$
where $\FFN$ is a fully connected feed-forward network, which consists of two linear transformations with a ReLU activation in between:
$$\FFN(\tilde{\Vh}_v^{(n)}) = \MW_2\max(\mathbf{0}, \MW_1\tilde{\Vh}_v^{(n)} + \Vb_1) + \Vb_2$$
where $\MW_1$, $\MW_2$, $\Vb_1$ and $\Vb_2$ are learnable parameters, and $\tilde{\Vh}_v^{(n)} \in \mathbb{R}^d$ is the $i^\text{th}$ row of $\tilde{\MH}_v^{(n)}$ representing $i^\text{th}$ vertex node in the graph.

\section{\dsname{} Dataset}

\begin{figure*}[ht]
  \centering
  \includegraphics[width=0.9\linewidth]{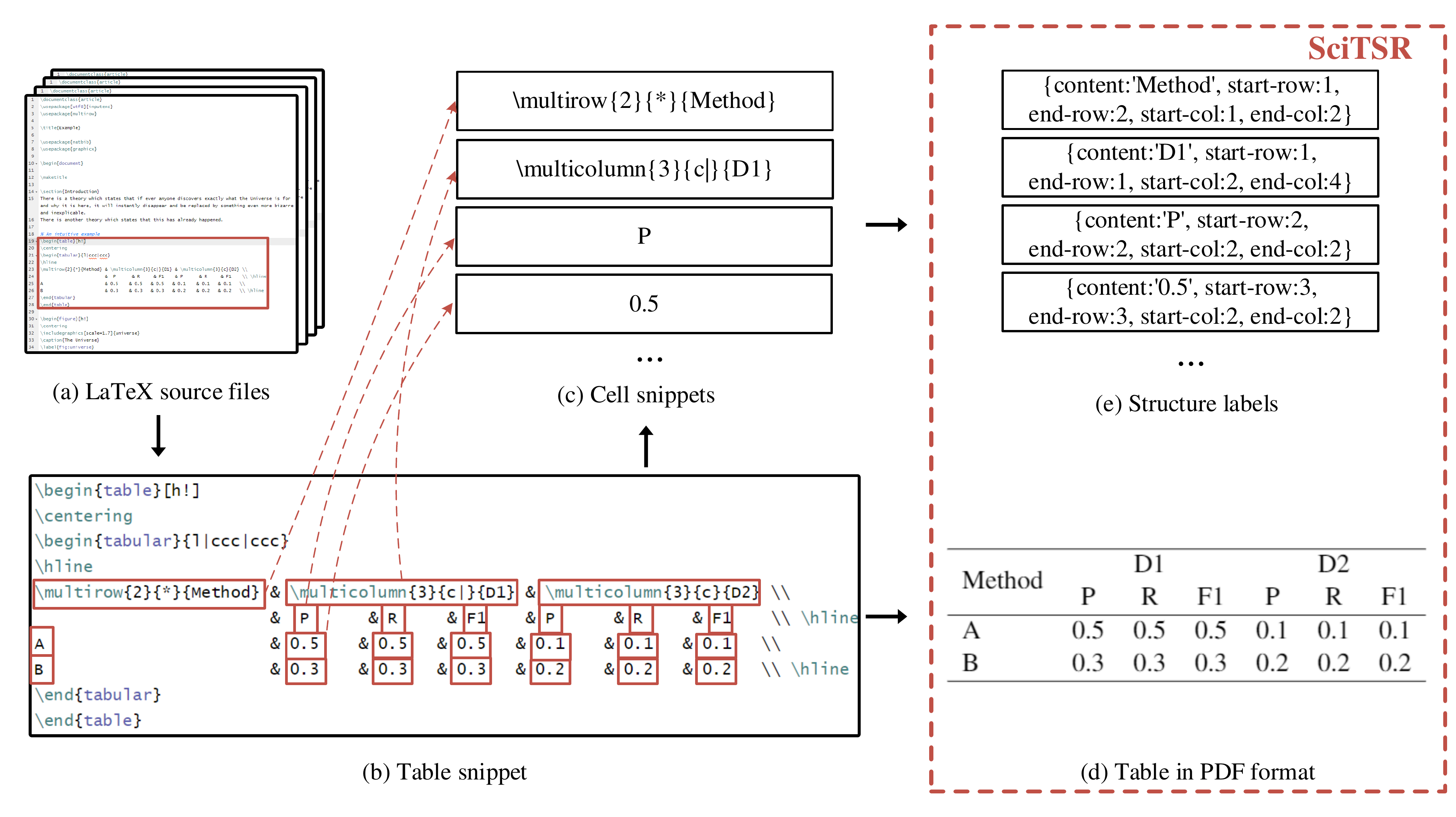}
  \caption{The construction pipeline of \dsname{} dataset.}
  \label{fig:latex}
\end{figure*}

We construct a large-scale table structure recognition dataset, named \dsname{}, which contains 15,000 tables in PDF format and their corresponding high quality structure labels obtained from LaTeX source files.
In this section, we will introduce the construction details and statistics of our \dsname{} dataset.


\subsection{Construction}

The construction pipeline of \dsname{} dataset is shown in Figure \ref{fig:latex}. 
We first crawl LaTeX source files from arXiv\footnote{\url{https://arxiv.org/}}, where there are a large number of papers in LaTeX format.
Then we extract all the table snippets (Figure \ref{fig:latex} (b)) from the LaTeX source files by a regular expression. A table snippet is a LaTeX code snippet used to present tables, which usually begins with `\verb!\begin{table}!' command and ends with `\verb!\end{table}!' command.
After that, we compile each table snippet to an individual PDF file (Figure \ref{fig:latex} (d)).
To obtain the structure labels, we split each table snippet with `\verb!\\!' and `\verb!&!', thus we get a series of cell snippets (Figure \ref{fig:latex} (c)) and their corresponding coordinates (i.e., start row, end row, start column and end column).
As there are many kinds of commands, such as `\verb!\textbf{}!' and `\verb!\alpha!', in the code, the cell snippets are not always same as the actual text in PDF. So we compile the cell snippets into PDF files, and then extract the real text from them.
Note that there are several commands like `\verb!\multirow{}!' or `\verb!\multicolumn{}!' in the cell snippets, which means these cells are spanning cells. Therefore, we re-compute their coordinates by parsing the `\verb!\multirow{}!' and `\verb!\multicolumn{}!' commands.
By now, we get all the cell contents and their coordinates, namely structure labels, and we dumped them in JSON format (Figure \ref{fig:latex} (e)).

\subsection{Statistics}

\begin{table}[t]
\centering
\begin{tabular}{lrr}
\hline
                      & Train   & Test  \\ \hline
\#tables              & 12,000  & 3,000 \\
\#complicated tables  & 2,885    & 716   \\
Ratio of complicated tables  & 24.0\%  & 23.9\% \\
\hline
Avg. \#rows / table   & 9.29    & 9.31  \\
Avg. \#columns / table   & 5.21    & 5.18  \\
Avg. \#cells / table  & 47.74   & 48.80 \\
\hline
\end{tabular}
\caption{The statistics of \dsname{} dataset.}
\label{tab:ds-stat-1}
\end{table}

\begin{table}[t]
\centering
\begin{tabular}{l|rr}
\hline
\multicolumn{1}{l|}{Dataset} & Train  & Test  \\ \hline
ICDAR-2013                   & 0      & 156  \\
\dsname{}                          & 12,000 & 3,000 \\
\dshard{}                    & -      & 716  \\ \hline
\end{tabular}
\caption{Number of tables in ICDAR-2013 and our proposed \dsname{} dataset.}
\label{table:dataset}
\end{table}

\begin{table*}[th]
\centering
\begin{tabular}{l|ccc|ccc|ccc}
\hline
\multirow{2}{*}{Method} & \multicolumn{3}{c|}{ICDAR-2013}                  & \multicolumn{3}{c|}{\dsname{}}                         & \multicolumn{3}{c}{\dshard{}}                    \\
                        & Precision      & Recall         & F1             & Precision      & Recall         & F1             & Precision      & Recall         & F1             \\ \hline
Tabby                   & 0.789          & 0.845          & 0.816          & 0.914          & 0.910          & 0.912          & 0.869          & 0.841          & 0.855          \\
DeepDeSRT               & 0.573          & 0.564          & 0.568          & 0.898          & 0.897          & 0.897          & 0.811          & 0.813          & 0.812          \\
Adobe                   & -              & -              & -              & 0.829          & 0.796          & 0.812          & 0.796          & 0.737          & 0.765          \\
\modelname{}               & \textbf{0.819} & \textbf{0.855} & \textbf{0.837} & \textbf{0.936} & \textbf{0.931} & \textbf{0.934} & \textbf{0.943} & \textbf{0.925} & \textbf{0.934} \\ \hline
\end{tabular}
\caption{Macro-averaged experiment results on ICDAR-2013, \dsname{} and \dshard{} dataset. }
\label{table:result-macro}
\end{table*}

\begin{table*}[th]
\centering
\begin{tabular}{l|ccc|ccc|ccc}
\hline
\multirow{2}{*}{Method} & \multicolumn{3}{c|}{ICDAR-2013}                  & \multicolumn{3}{c|}{\dsname{}}                         & \multicolumn{3}{c}{\dshard{}}                    \\
                        & Precision      & Recall         & F1             & Precision      & Recall         & F1             & Precision      & Recall         & F1             \\ \hline
Tabby                   & 0.846          & \textbf{0.862} & 0.854          & 0.926          & 0.920          & 0.921          & 0.892          & 0.872          & 0.882          \\
DeepDeSRT               & 0.632          & 0.617          & 0.615          & 0.906          & 0.887          & 0.890          & 0.863          & 0.831          & 0.846          \\
Adobe                   & -              & -              & -              & 0.930          & 0.784          & 0.851          & 0.901          & 0.717          & 0.798          \\
\modelname{}              & \textbf{0.885} & 0.860          & \textbf{0.872} & \textbf{0.959} & \textbf{0.948} & \textbf{0.953} & \textbf{0.964} & \textbf{0.945} & \textbf{0.955} \\ \hline
\end{tabular}
\caption{Micro-averaged experiment results on ICDAR-2013, \dsname{} and \dshard{} dataset. }
\label{table:result-micro}
\end{table*}

The statistics of the \dsname{} dataset are shown in Table \ref{tab:ds-stat-1}. There are 15,000 tables as well as their corresponding structure labels in total, and we split 12,000 for training and 3,000 for test. There are averagely about 9 rows, 5 columns and 48 cells in a table.
Particularly, we focus on the complicated tables, which have at least one spanning cell in them.
There are 2,885 and 716 complicated tables in training and test set, about 24.0\% and 23.9\%, respectively.
That means the majority of tables in \dsname{} dataset are still simple grid-like tables.
Furthermore, to reflect the model's ability of recognizing complicated tables, we extract all the 716 complicated tables from the test set as a test subset, called \dshard{}. 

\section{Experiment}

\subsection{Dataset}

We evaluate our model and baselines on our \dsname{} dataset 
and the widely used ICDAR-2013 dataset \cite{gobel2013icdar}.
Both datasets provide tables in PDF format and the corresponding structure labels, which contains contents and coordinates (i.e., the number of start row, end row, start column and end column) of cells.
It should be noted that ICDAR-2013 dataset doesn't have a training set, but only provides a small test set.
While our \dsname{} dataset provides a large number of tables for both train and test set.
Statistics of these datasets are listed in Table \ref{table:dataset}.

\subsection{Metrics}

We employ the widely used evaluation procedure presented by \citet{gobel2012methodology}, which is also used in ICDAR 2013 table competition.
Specifically, it first converts a table to a list of horizontally and vertically adjacent relations between cells and their vertical and horizontal neighbors, and then make a comparison on relations extracted from output tables and ground truth by using precision and recall measures.
We first calculate these scores separately on each table, and then compute both macro- and micro-averaged scores as the final result.

\subsection{Baselines}

We compare our method with two state-of-the-art baselines and a commercial software:

\begin{itemize}
  \item \textbf{Tabby}: The Tabby \cite{shigarov2016configurable} is a rule-based tool for extracting tables from PDF documents, and we use their open-source implementation for experiments\footnote{\url{http://github.com/cellsrg/tabbypdf/}}.
  \item \textbf{DeepDeSRT}: DeepDeSRT \cite{schreiber2017deepdesrt} is a data-driven method that utilizes a semantic segmentation model to recognize the table structure as a set of segmented rows and columns. We implemented this model because there is no available code for this method.
  \item \textbf{Adobe}: We use the Adobe Acrobat DC SDK\footnote{\url{https://www.adobe.com/devnet/acrobat/sdk/eula.html}} to extract tables to HTML format, and then parse these files to obtain structure labels. For the sake of simplicity, we denote it as Adobe.
\end{itemize}
The tag sequences generated by the image-to-text model \cite{li2019tablebank} don't provide column coordinates. They cannot be used to restore the origin tables. So we don't use it as a baseline in our experiment.

\begin{table*}[th]
\centering
\begin{tabular}{l|ccc|ccc}
\hline
\multirow{2}{*}{Method} & \multicolumn{3}{c|}{Macro}                      & \multicolumn{3}{c}{Micro}                      \\
                        & Precision      & Recall         & F1             & Precision      & Recall         & F1             \\ \hline
Tabby                   & 0.363          & 0.397          & 0.379          & 0.141          & 0.332          & 0.196          \\
DeepDeSRT               & -              & -              & -              & -              & -              & -              \\
Adobe                   & 0.480          & 0.490          & 0.485          & \textbf{0.647} & 0.468          & 0.543          \\
\modelname{}               & \textbf{0.711} & \textbf{0.696} & \textbf{0.703} & 0.630          & \textbf{0.620} & \textbf{0.625} \\ \hline
\end{tabular}
\caption{Experiment results on complicated tables in \dshard{} where adjacent relations among non-spanning cells are not considered. }
\label{table:result-hard}
\end{table*}

\subsection{Implementation Details}

We implement our model with PyTorch 0.4.1, and train a 4-block \modelname{}
with $d=64$ for both edge-to-vertex and vertex-to-edge attention blocks.
It's trained to minimize the cross-entropy on the labeled edges using the Adam \cite{kingma2014adam} 
optimizer with an initial learning rate of 0.0005.
Because most of the edges constructed by $K$NN is labeled as \textit{``no relation''},
we set a manual rescaling weight of 0.2 for \textit{``no relation''} and 1.0 for the \textit{``vertical''} and \textit{``horizontal''} relations.
Besides, when constructing edges, $K$ is set as $20$ to cover most real edges.
We also use a $L_2$ weight decay with $\lambda = 0.0001$ on parameters and dropout
\cite{srivastava2014dropout} with $p=0.4$
to the output of each sub-layer before it is added to the sub-layer input and normalized
to prevent over-fitting.
During training, we utilize a batch size of 1 graph for 15 epochs on Intel Xeon CPUs,
and each epoch takes about 20 minutes for 12,000 tables in total.
Because our proposed \modelname{} cannot directly take PDF files as input, we compute a matching between pre-processed cells from input PDFs and ground
truth cells generated by LaTeX documents to label edges as training data.

\subsection{Result and Discussion}

\paragraph{Overall results} 

Our main results are shown in Table \ref{table:result-macro} and \ref{table:result-micro}, where the results are presented by macro- and micro-averages scores, respectively.
From the tables, it can be observed that: (1) Our model outperforms state-of-the-art baselines on all datasets. On both ICDAR-2013 and \dsname{} dataset, F1 scores of our method are at least 2\% higher than baselines.
While in \dshard{}, our method outperforms other methods at least 7\% in F1 scores, providing a significant improvement. 
(2) Our model shows a strong generalization ability.
Both DeepDeSRT and our model are trained on the \dsname{} training set and tested on ICDAR-2013 dataset because it doesn't provide a training set.
With the same training set, we can see that DeepDeSRT suffers from a big drawback on the test set of ICDAR-2013, because there is a domain difference between ICDAR-2013 dataset and our \dsname{} dataset.
In contrast, our model still achieves the best results comparing to the other three baselines, which demonstrates the strong generalization ability of our model.
It also suggests that directly taking images as input makes the model sensitive to fonts or styles of tables, and finally fails to generalize on tables with unseen fonts or styles.

\begin{figure*}[th]
\begin{center} 
\includegraphics[width=0.90\linewidth]{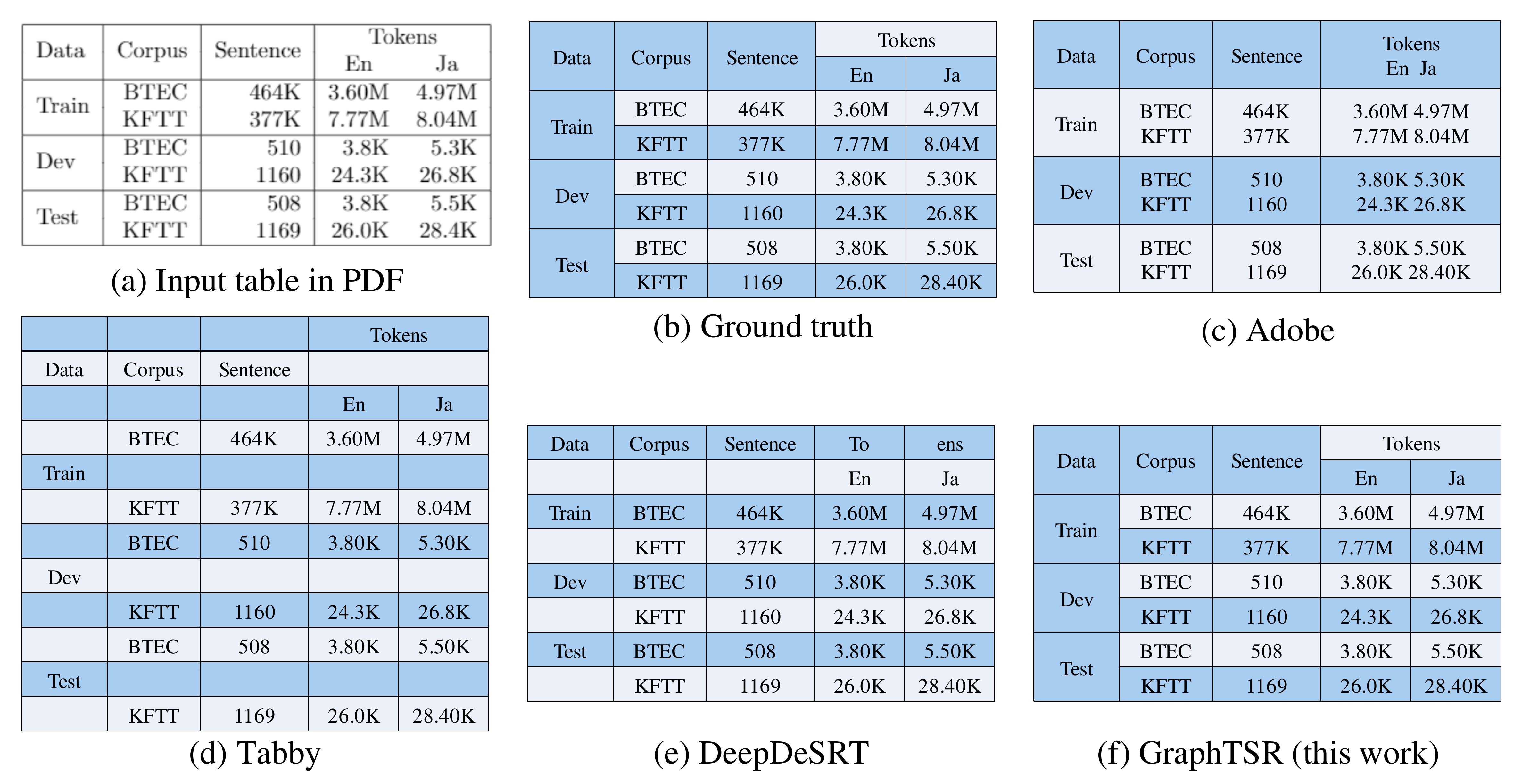}
\caption{A sample from results on \dshard{}. Cells are marked with different colors to distinguish from each other.} 
\label{fig:case}
\end{center} 
\end{figure*}

\begin{figure}[th]
\begin{center} 
\includegraphics[width=1.0\linewidth]{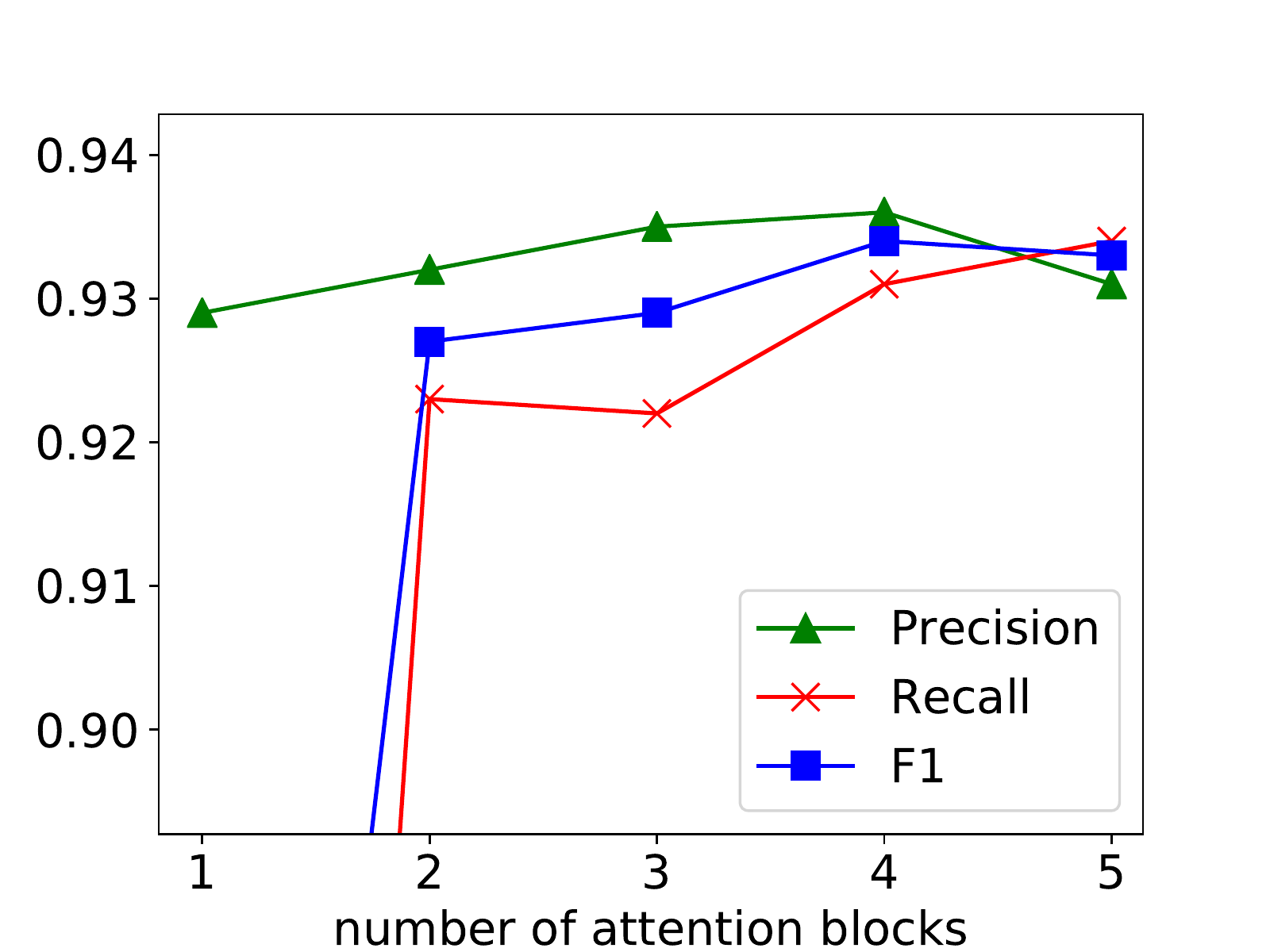}
\caption{Performance of our model with various numbers of graph attention blocks on \dsname{} dataset, and the scores are macro-averaged. } 
\label{fig:N}
\end{center} 
\end{figure}

\paragraph{Results on complicated tables}
To evaluate the power of our model for complicated tables, we conduct experiments on \dshard{} dataset, and the results are shown in Table \ref{table:result-macro} and \ref{table:result-micro}.
Comparing to the results on \dsname{} dataset, we observe that all the baselines have at least a 4\% performance drop off on \dshard{} while our method remains a high level of F1 scores, which indicates that our model is better able to capture structures of complicated tables.
Though the \dshard{} test set only contains complicated tables, the dominant type of cells is still the non-spanning cells.
The results on \dshard{} cannot perfectly show the power of our model on complicated tables.
Therefore, we perform additional experiments that on only spanning cells, which means relations between non-spanning cells won't be considered in the evaluation.
The experiment results are presented in Table \ref{table:result-hard}, where all the methods have different degrees of decline in performance, especially Tabby, which reaches a decrease of micro-F1 to 72.5\%. However, our method consistently outperforms the baselines greatly.
Note that as the output of DeepDeSRT is always grid-like tables without spanning cells, it cannot recognize any spanning cells.

\paragraph{Case study}
We collect the outputs of these methods on \dshard{} test set, and perform a case study to analyze the advantages of our model on complicated tables.
Figure \ref{fig:case} shows an example from \dshard{}, which is a table presented in a scientific paper\footnote{\url{https://aclweb.org/anthology/D16-1162}}.
We compare the ground truth table structures and the recognized table structures by different methods, where cells are marked with different colors to distinguish from each other. From the results, it can be found that our model recognizes the internal structure of the table correctly while other methods have different degrees of mistakes. In Figure \ref{fig:case} (c), Adobe simply recognizes the structure as a grid-like table where contents in the same ruling box are incorrectly merged into a single cell. While in Figure \ref{fig:case} (d), the headers and their corresponding body cells are placed into different rows because of the small horizontal overlap (i.e., \textit{Train} and \textit{BTEC}).
In Figure \ref{fig:case} (e), limited to the design of the model, the DeepDeSRT can only recognize a table as a grid-like structure, so all the spanning cells are split into non-spanning cells.

\paragraph{Impact of the number of attention blocks} To better understand the impact of the number of graph attention blocks $N$ in the \modelname{}, we perform a study by changing $N$. The results is illustrated in Figure \ref{fig:N}. It can be observed that the performance of our model improves as $N$ increases. Moreover, we find that $N$ has a great impact on the recall scores. It suggests that if $N$ is set to be too small, nodes in the graph can only access limited surrounding nodes, resulting in the conservative prediction of the model. In other words, when $N$ is small, the model tends to predict \textit{``no relation''} between two cells.

\section{Conclusion}

In this paper, we propose a novel graph neural model for complicated table structure recognition in PDF files. Moreover, we release a large-scale dataset for table structure recognition in PDF files, which contains 15,000 tables and their corresponding structure labels. Extensive experiments show the power of our model on complicated tables.

\bibliography{ref}
\bibliographystyle{acl_natbib}

\end{document}